\documentclass{aa-bis}

\newcommand\dosingle[1]{#1}  \newcommand\dodouble[1]{ } 

\newcommand\nice[1]{#1}    \newcommand\subm[1]{}   
\subm{ \renewcommand\dosingle[1]{ }   \renewcommand\dodouble[1]{#1} }

\dodouble{ \documentclass[referee]{aa} }
\usepackage{epsf1990}  

\usepackage{url} 

\newcommand\zzz[2]{#2} 

\newcommand\SSS{Sect.~}

\nice{ \newcommand\mycaptionfont{\protect\footnotesize} }

\nice{ \textheight 240mm \headheight 0pt \topmargin -2.1cm }

\newcommand\centreline{\centerline}
\newcommand\gtapprox{\,\lower.6ex\hbox{$\buildrel >\over \sim$} \, }
\newcommand\ltapprox{\,\lower.6ex\hbox{$\buildrel <\over \sim$} \, }
\newcommand\propapprox{\,\lower.6ex\hbox{$\buildrel \propto\over \sim$} \, }

\newcommand\arcs{\ifmmode {'' }\else $'' $\fi}     
\newcommand\arcm{\ifmmode {' }\else $' $\fi}       
\newcommand\ddeg{\ifmmode^\circ\else$^\circ$\fi}    

\newcommand\frtoday{Le\space\number\day\space\ifcase\month\or
  janvier\or f\'evrier\or mars\or avril\or mai\or juin\or
  juillet\or ao\^ut\or septembre\or octobre\or novembre\or 
d\'ecembre\fi\space \number\year}

\newcommand\ddd{\mbox{d}}

\title{A Solution to the Isolatitude,  Equi-area, Hierarchical 
Pixel-Coordinate System
}

\author{Boudewijn F. Roukema,
Bartosz Lew
}

\institute{Toru\'n Centre for Astronomy, N. Copernicus University,
ul. Gagarina 11, PL-87-100 Toru\'n, Poland 
}

\date{\frtoday}

\titlerunning{Isolatitude, Equi-area, Hierarchical Pixelisation
}
\authorrunning{Roukema et al.}

\begin{document}

\abstract{
\protect\nocite{Healpix99}{G\'orski} {et~al.} (1999b) have earlier presented 
the outline 
of a pixelisation-to-spherical-coordinate transformation scheme
which simultaneously satisfies three 
properties which are especially useful for 
rapid analyses of maps on a sphere:
(i) equal spacing of pixels along lines of constant latitude, 
(ii) equal pixel `areas' (solid angles) 
and (iii) hierarchical scaling with increasing
numbers of pixels. Their outline is based on the division of the sphere
into twelve regions covering equal solid angles, which are hierarchically
subdivided in a way compatible with these three criteria.
In this paper, a complete derivation of this 
scheme is presented, including, in particular,
(1) the angle $\theta^*$ defining the limit 
between polar and equatorial regions, and
(2) the transformations from the unit interval $[0,1] \wedge [0,1]$ 
to spherical coordinates in a polar region.
}

\maketitle

\dodouble{ \clearpage } 


\newcommand\ffifjeq{
\begin{figure}
\centering 
\nice{ \centreline{\epsfxsize=8cm
\zzz{\epsfbox[0 0 493 343]{"`gunzip -c fifjeq.eps.gz"}}
{\epsfbox[0 0 493 383]{"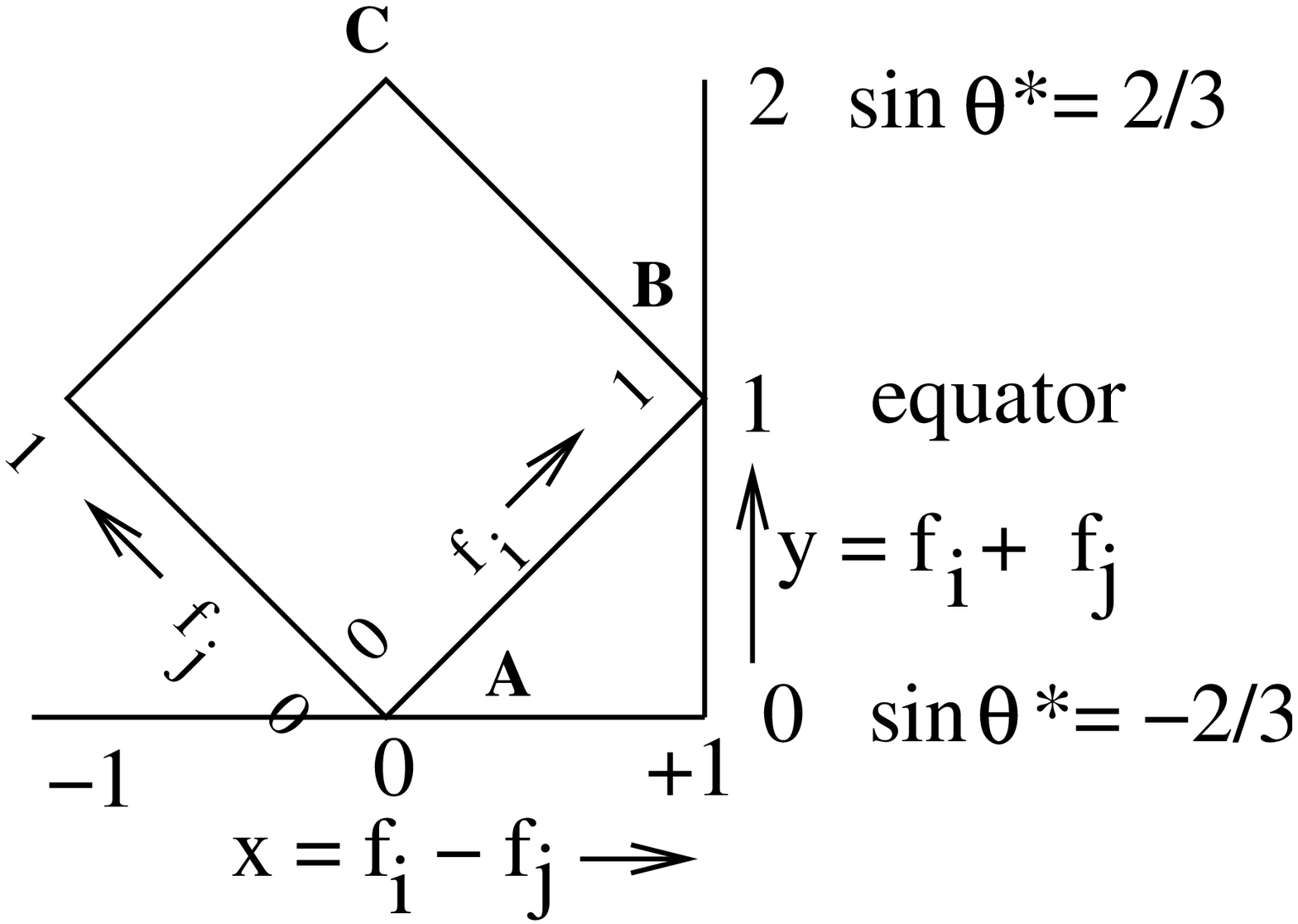"}}  } }
\caption[]{ \mycaptionfont
Equatorial region: 
unit interval $[0,1] \wedge [0,1]$ parametrised by fractions 
$f_i, f_j$ and linearly transformed to $x \equiv f_i - f_j$ 
and $y \equiv f_i + f_j$. This can be any of the regions 5, 6, 7, 8; 
without loss of generality, the discussion in \SSS\protect\ref{s-fifjeq} is
presented for region 5.
}
\label{f-fifjeq}
\end{figure} 
} 

\newcommand\ffifj{
\begin{figure}
\centering 
\nice{ \centreline{\epsfxsize=8cm
\zzz{\epsfbox[0 0 493 343]{"`gunzip -c fifj.eps.gz"}}
{\epsfbox[0 0 493 383]{"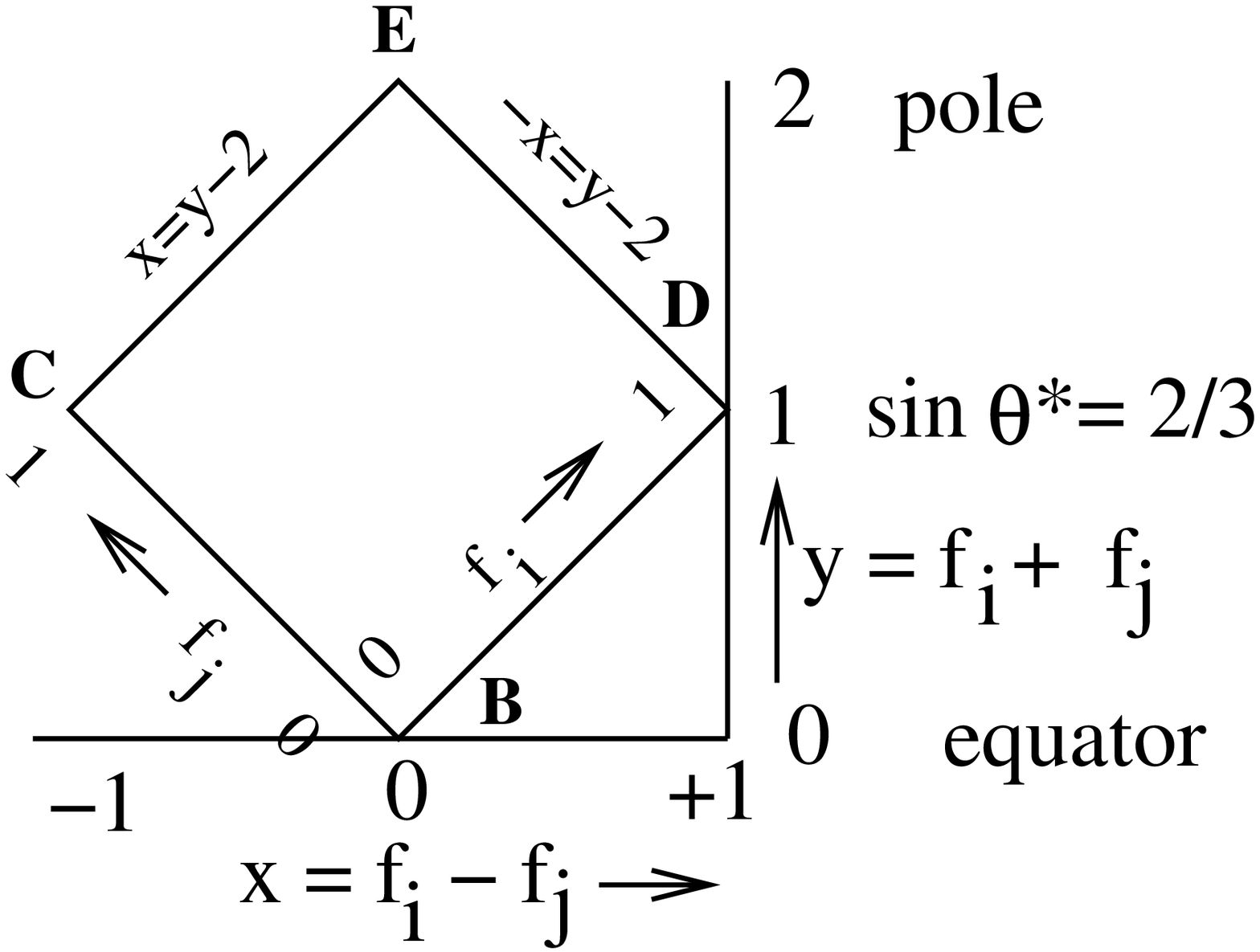"}}  } }
\caption[]{ \mycaptionfont
Polar region: unit interval $[0,1] \wedge [0,1]$ parametrised by fractions 
$f_i, f_j$ and linearly transformed to $x \equiv f_i - f_j$ 
and $y \equiv f_i + f_j$. This can be any of the regions 1, 2, 3, 4, 
9, 10, 11, 12; 
without loss of generality, the discussion in \SSS\protect\ref{s-fifjpol} is
presented for region 1.
}
\label{f-fifj}
\end{figure} 
} 

\newcommand\twopithree{
\begin{figure}
\centering 
\nice{ \centreline{\epsfxsize=8cm
\zzz{\epsfbox[0 0 686 333]{"`gunzip -c twopi3.eps.gz"}}
{\epsfbox[20 0 760 350]{"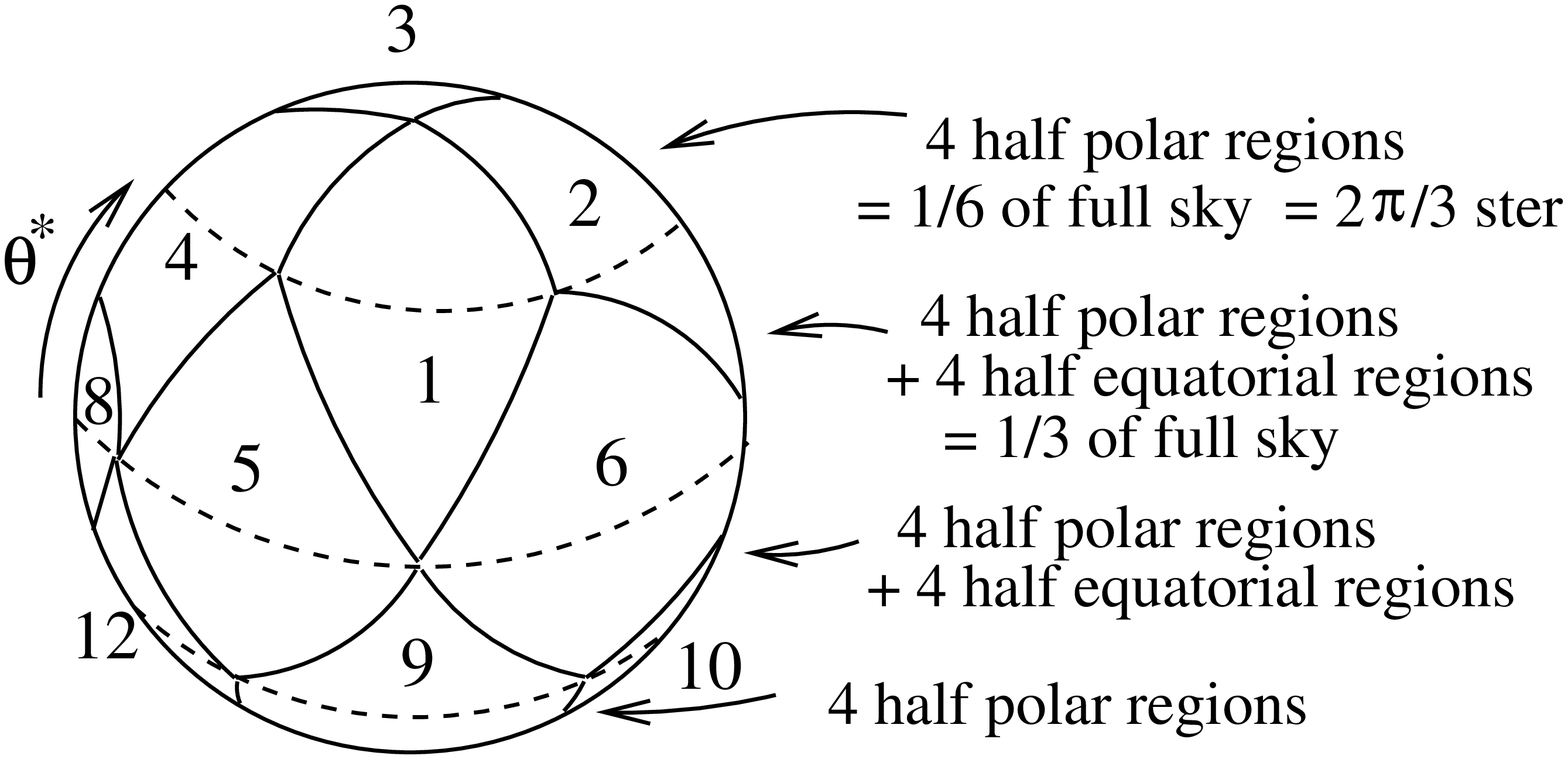"}}  } }
\caption[]{ \mycaptionfont
Schematic diagram of the 12 regions, or equivalently, 24 half regions,
showing how the isolatitude requirement and the equal pixel
area requirement together require that the
solid angle from the line of constant latitude $\theta^*$ to the pole
is $2\pi/3$~ster.
}
\label{f-twopithree}
\end{figure} 
} 

\newcommand\twelveregions{
\begin{figure*}
\centering 
\nice{ \centreline{\epsfxsize=15cm
\zzz{\epsfbox[50 0 1131 690]{"`gunzip -c twelveregions.eps.gz"}}
{\epsfbox[50 0 1131 690]{"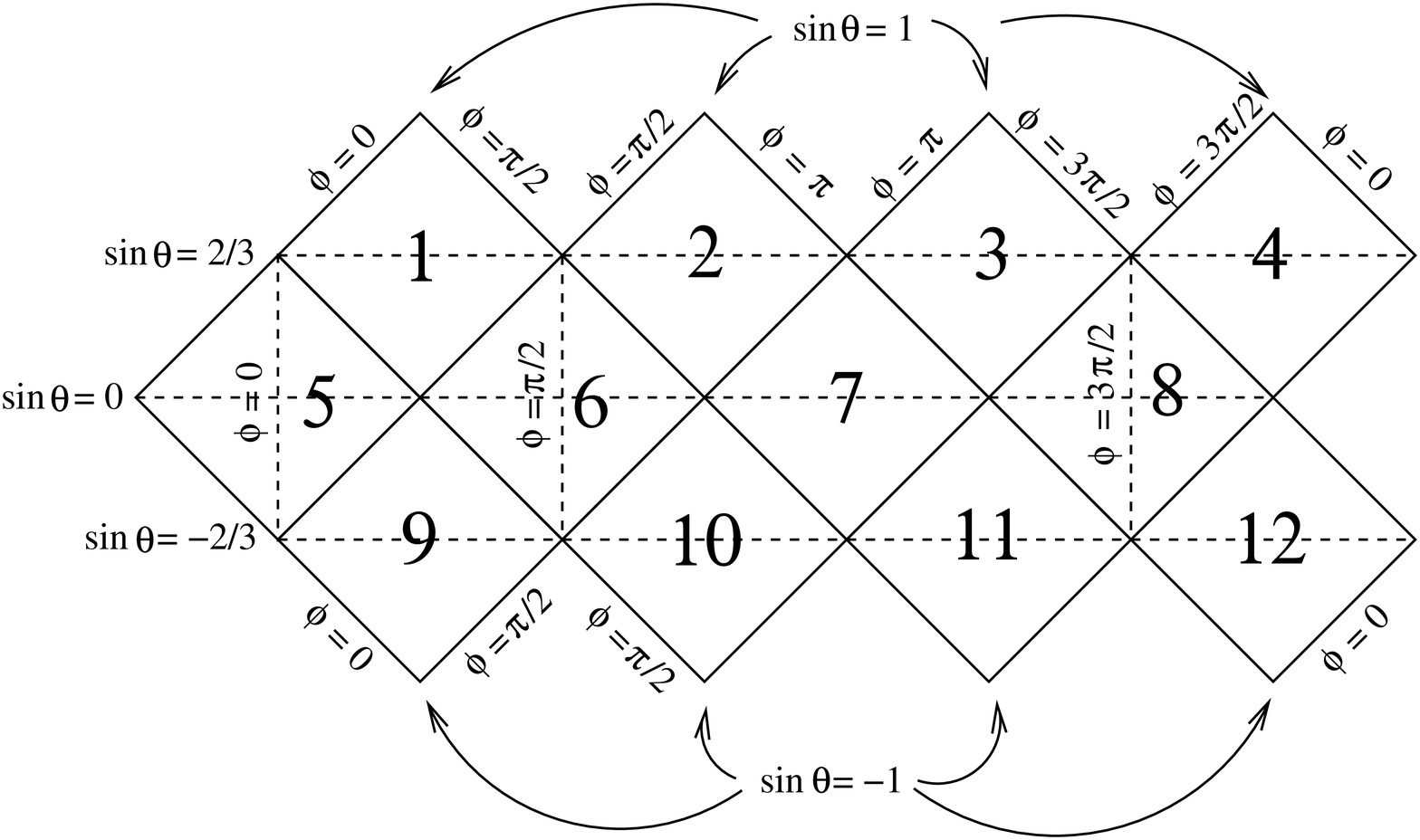"}}  } }
\caption[]{ \mycaptionfont
Ordering of the 12 regions, shown in a projection where the full
sky, $4\pi$~ster, is shown as 12 square unit intervals. 
Each pole is represented as a set of four distinct
points in this projection, indicated by arrows.
The derivation of the straight line (in this projection) dividing 
the ``polar'' regions into ``polar'' and ``equatorial'' parts,
labelled $\sin \theta = 2/3$, is 
derived in Eq.~(\protect\ref{e-sintheta}).
}
\label{f-twelveregions}
\end{figure*} 
} 

\newcommand\tbcequat{
\begin{table}
\caption{Vertex conversions for an equatorial region. 
Equations (\protect\ref{e-defnxy}), (\protect\ref{e-defnphipi}) and
(\protect\ref{e-defns}) define $x, y, \phi_\pi$ and $s$. 
See eqns (\protect\ref{e-phieqval}), (\protect\ref{e-phieqpolval}) 
and (\protect\ref{e-sintheta}) for the values of 
$\phi_\pi$ and $s$.
\label{t-bcequat}}
\begin{equation}
\begin{array}{c c c c c} \hline 
\rule[-0.4ex]{0ex}{2.5ex}
\mbox{label} & x & y & \phi_\pi & s \\ \hline 
   A &     0 & 0   & 0 & -2/3  \\
   B &     1 &  1  & 1/4  & 0 \\
   C &     0 &  2 & 0 &   2/3 \\
\hline
\end{array}
\label{e-bcequat}
\end{equation}
\end{table}
}  

\newcommand\tbcpolar{
\begin{table}
\caption{Vertex conversions (B to E) 
and a boundary condition (F) for a polar region. 
Equations (\protect\ref{e-defnxy}), (\protect\ref{e-defnphipi}) and
(\protect\ref{e-defns}) define $x, y, \phi_\pi$ and $s$. 
See eqns (\protect\ref{e-phipolval}), (\protect\ref{e-phieqpolval}) 
and (\protect\ref{e-sintheta}) 
for the values of 
$\phi_\pi$ and $s$.
\label{t-bcpolar}}
\begin{equation}
\begin{array}{c c c c c} \hline 
\rule[-0.4ex]{0ex}{2.5ex}
\mbox{label}  & x & y &  \phi_\pi & s \\ \hline 
   B &      0 & 0  & 1/4 & 0 \\
   C &     -1 &  1  & 0 & 2/3 \\
   D &     +1 &  1  & 1/2 & 2/3 \\
   E &     0 &  2   & 
\; 
0 \le  \lim_{(x,y) \rightarrow \mbox{\small E}} \phi_\pi \le 1/2 
& 1 
\rule[-1.8ex]{0ex}{2.0ex}
\\
  \cline{2-5}
\rule[-1.2ex]{0ex}{3.9ex}
F & \multicolumn{4}{c}{ \phi_\pi(-x=y-2) = \phi_\pi(x=y-2) + 1/2 } \\
\hline
\end{array}
\label{e-bcpolar}
\end{equation}
\end{table}
}  


\section{Introduction}

The most commonly required analyses of microwave background data
require spherical harmonic analyses of temperature fluctuation maps.
Given increasingly large numbers of pixels, \nocite{Healpix98,Healpix99}{G\'orski}, {Hivon}, \&  {Wandelt} (1999a); {G\'orski} {et~al.} (1999b) 
pointed
out that an ideal pixelisation-to-spherical-coordinate transformation
scheme would satisfy the following requirements:
\begin{list}{(\roman{enumi})}{\usecounter{enumi}}
\item an isolatitude requirement, i.e. 
pixels should be equally spaced in latitude, 
\item pixel solid angles (hereafter, `areas') should be equal, and
\item the system should scale hierarchically with increasing
numbers of pixels, preferably according to some power of two, in order
to maximally exploit the binary nature of present computer technology.
\end{list}

\nocite{Healpix99}{G\'orski} {et~al.} (1999b) present the outline of a system
satisfying these requirements. The idea of dividing the sphere into
12 regions --- four equatorial regions and eight polar regions --- 
which are mapped to or from 
the unit interval $[0,1] \wedge [0,1]$, is presented there.

However, the full solution is not derived in \nocite{Healpix99}{G\'orski} {et~al.} (1999b).

In this paper, firstly, in \SSS\ref{s-thetastar}, the angle $\theta^*$
defining the limit between polar and equatorial regions of this scheme
is derived.

As is described in \nocite{Healpix99}{G\'orski} {et~al.} (1999b), the binary bits of fractions 
with the square unit interval $[0,1] \wedge [0,1]$ are used 
for pixel ordering within a region. For completeness, we remind the reader 
of this ordering in \SSS\ref{s-hierarchical} and express it algebraically.

In order to have the complete transformation from pixel numbers to 
spherical angles,  
the 
transformation between the fractional position $(f_i,f_j)$ 
in the square unit interval 
and a pair of
spherical angles $(\phi, \theta)$ is needed.
In \SSS\ref{s-fifjeq}, this transformation is derived for equatorial
regions, and in \SSS\ref{s-fifjpol}, the transformation is derived for
polar regions.

\twopithree

\twelveregions

The convention chosen here for spherical coordinates 
is to have a longitude coordinate 
\begin{equation}
\phi \in [0,2\pi]
\end{equation}
and a latitude coordinate 
\begin{equation}
\theta \in [-\pi/2, \pi/2].
\end{equation}

For clarity of presentation, it is also useful to define 
\begin{equation}
\phi_\pi \equiv \phi/\pi 
\label{e-defnphipi}
\end{equation}
and 
\begin{equation}
s \equiv \sin \theta.
\label{e-defns}
\end{equation}

\section{Calculations}

\subsection{The boundary between equatorial and polar regions: $\theta^*$} 
\label{s-thetastar}

Fig.~\ref{f-twopithree} (cf. fig.~2 of \nocite{Healpix99}{G\'orski} {et~al.} (1999b)) 
and Fig.~\ref{f-twelveregions} show the 
overall geometry of the 12 regions, including an ordering, though any
other ordering could easily be chosen.

In order for the 12 regions to have equal pixel areas, 
the coordinates of the four vertices along the equator, of the equatorial regions,
must be spaced by an interval of $\Delta \phi = \pi/2$, i.e. they are
\begin{equation}
\phi_\pi = 1/4,\; 3/4,\; 5/4,\; 7/4, \theta = 0.
\label{e-phieqval}
\end{equation}
The two polar vertices are
\begin{equation}
\phi_\pi \; \mbox{undefined},\; \theta = -\pi/2, +\pi/2.
\label{e-phipolval}
\end{equation}

Finding the coordinates of the remaining eight vertices, which separate
equatorial and polar regions, requires an integration.

Define $\theta^*$ to be the angle from the equator 
to the curve of constant latitude separating the equatorial regions
from the ``polar'' half of the polar regions, as shown in 
Fig.~\ref{f-twopithree}. (The ``polar'' regions are, in fact, divided between
a ``polar'' half which is polar and an ``equatorial'' half which 
uses a transformation similar to that of the ``equatorial'' regions. This
is explained below in \SSS\ref{s-fifjpol}.)

Since the system is an isolatitude, equal pixel area system, 
the entire solid angle from a curve of constant latitude 
$\theta^*$ to the pole must be half of the
total solid angle represented by 
four regions, i.e. half of one third of the full sky, i.e. 
one sixth of $4\pi$ steradians, i.e. $2\pi/3$~ster. This is illustrated
in Fig.~\ref{f-twopithree}.

Hence, 
\begin{eqnarray}
\int_0^{2\pi} \ddd \phi \int_{\theta^*}^{\pi/2} \cos \theta \; \ddd \theta
&=& 2\pi/3, 
\nonumber
\end{eqnarray}
\begin{eqnarray}
\Rightarrow 2\pi \left[ \sin \theta \right]_{\theta^*}^{\pi/2} &=&  2\pi/3
  \nonumber
\end{eqnarray}
\begin{eqnarray}
\Rightarrow 2\pi (1 - \sin \theta^*) &=&  2\pi/3  
  \nonumber
\end{eqnarray}
\begin{eqnarray}
\Rightarrow \sin \theta^* &=&  2/3
\label{e-sintheta}
\end{eqnarray}
\begin{eqnarray}
\Rightarrow \theta^* &=&  \sin^{-1} 2/3 \approx 41.8\ddeg.
\end{eqnarray}

Hence, in Fig.~\ref{f-twelveregions}, the curve of constant latitude
dividing each polar region into two equal halves is labelled 
$\sin \theta = \pm 2/3$.

Note that $\theta^*$ is {\em not} $\pi/4$, even though a casual
glance at Fig.~\ref{f-twopithree} and casual intuition might 
suggest this.

Hence, the eight vertices separating equatorial and polar regions
are 
\begin{eqnarray}
&(\phi_\pi = 0,\; 1/2,\; 1,\; 3/2, \; \theta = \theta^*),&  \nonumber \\
&(\phi_\pi = 0,\; 1/2,\; 1,\; 3/2, \; \theta = -\theta^*).&  
\label{e-phieqpolval}
\end{eqnarray}

\ffifjeq

\ffifj

\subsection{Hierarchical ordering of pixels}
\label{s-hierarchical}

Requirement (iii), the hierarchical ordering, is a binary ordering
within the (square) unit interval, and an ordering of the twelve
unit intervals. This ordering is referred to as
`nested' by \nocite{Healpix99}{G\'orski} {et~al.} (1999b), and is the ordering used in the data
files of the first year WMAP data \nocite{WMAPbasic}({Bennett} {et~al.} 2003), where $\phi$
and $\theta$ are galactic longitude and latitude respectively. It is 
shown for two different resolution levels in the lower half
of fig.~3 of \nocite{Healpix99}{G\'orski} {et~al.} (1999b).

\subsubsection{Binary ordering of pixels within the square unit interval}
\label{s-unitinterval}

Figs~\ref{f-fifjeq}, \ref{f-fifj} show
the definitions of 
$f_i, f_j$ as fractions in the unit interval in an equatorial
and a polar region respectively.

Let us write $f_i, f_j$ as binary fractions, rounded to the nearest
pixel centre, to some precision $n \ge 1$, corresponding 
to linear pixel sizes of {\em approximately},
but not exactly, ${\pi/2 \mbox{~rad} \over 2^n}$, or pixel {\em areas}
of {\em exactly} ${\pi/3 \mbox{~ster} \over 4^n}$:
\begin{equation}
0 \le f_i = 0.b_{2n-1}...b_3b_1 1, f_j= 0.b_{2n}...b_4b_2 1 \le 1 
\label{e-binarypixel}
\end{equation}
where $b_i \in \{0,1\}$. The last binary place is occupied by ``1'' 
in order to have a pixel centre.
Then the bits in $f_i$
and $f_j$ provide {\em alternating bits} for integer ordering of pixels within
the square unit interval. 

This mapping is sufficient for both equatorial and polar regions. 

This is shown in fig.~1 of \nocite{Healpix99}{G\'orski} {et~al.} (1999b).

For example, with just one binary digit, a region is divided into four
pixels. If we have binary representations
\begin{equation}
f_i = 0.b_1 1, f_j= 0.b_2 1,
\end{equation}
then the pixel centred 
at $(f_i,f_j)$ is the $(b_2b_1)_2$-th pixel within the
region. 

With two binary digits, a region is divided into sixteen
pixels. If we have binary representations
\begin{equation}
f_i = 0.b_3b_1 1, f_j= 0.b_4b_2 1,
\end{equation}
then the pixel centred at $(f_i,f_j)$ is the 
$(b_4b_3b_2b_1)_2$-th pixel within the
region.

More generally, with $n \ge 1$ binary digits, the pixel at $(f_i,f_j)$ 
\begin{equation}
f_i = 0.b_{2n-1}...b_3b_1 1, f_j= 0.b_{2n}...b_4b_2 1,
\end{equation}
is the $(b_{2n}...b_2b_1)_2$-th pixel (again in binary) in the region.

\subsubsection{Ordering of the twelve square unit intervals}
\label{s-twelveregions}

The ordering of the twelve regions shown in Fig.~\ref{f-twelveregions}
is that used in the first year WMAP data files. This combines with 
the ordering within a region.

That is, with $n \ge 1$ binary digits, the pixel centred at $(f_i,f_j)$ in 
the $k$-th region, 
with binary representations
\begin{equation}
f_i = 0.b_{2n-1}...b_3b_1 1, f_j= 0.b_{2n}...b_4b_2 1,
\end{equation}
is pixel number
\begin{equation}
\left[ (k-1)\; 2^{2n} \right]_{10} + \left[b_{2n}...b_2b_1\right]_2
\label{e-fullpixelnumber}
\end{equation}
where the first term is shown in decimal notation and the second
in binary notation.


\subsection{Transformations from the unit interval to $\phi,\theta$}
\label{e-unittophitheta}

For either an equatorial or polar region, 
shown in Figs~\ref{f-fifjeq}, \ref{f-fifj} respectively,
it is useful to make
the change of variables
\begin{eqnarray}
x &\equiv& f_i - f_j \nonumber \\ 
y &\equiv& f_i + f_j ,
\label{e-defnxy}
\end{eqnarray}
where $f_i, f_j$ are the fractions in the unit interval.

The requirements of
\begin{list}{(\roman{enumi})}{\usecounter{enumi}}
\item isolatitude pixel positions, and
\item equal pixel areas 
\end{list}
then require transformations from $(f_i, f_j)$ to 
$(\phi,\theta)$, or equivalently from $(x,y)$ to $(\phi,\theta)$, 
or even more simply from 
$(x,y)$ to $(\phi_\pi,s)$, as defined in 
eqs~(\ref{e-defnphipi}), (\ref{e-defns}).

Use of equal intervals in 
$s \equiv \sin \theta$ in latitude enables straightforward linear mappings
of the complete equatorial band (the regions labelled here as ``equatorial'',
plus half of each of the regions labelled here as ``polar''). This can
be thought of as stretching the latitude coordinate in order to compensate
for shorter lengths of curves of constant latitude.

However, this cannot function close to the poles, hence, a distinct mapping
is required for the polar regions.

\subsubsection{Equatorial regions: 
transformations from unit interval to $\phi,\theta$}
\label{s-fifjeq}

\tbcequat

Since 
\begin{equation}
\ddd (\sin \theta) = \cos \theta \; \ddd \theta,
\end{equation}
the mapping in 
any of the four equatorial regions is provided by a linear map from
$x$ to $\phi$ and from $y$ to $\sin \theta$. 

Without loss of generality (there are four equatorial regions in total),
it is sufficient to consider the region labelled ``5'' 
in Figs~\ref{f-twopithree},~\ref{f-twelveregions} to be 
the region which has $\phi=0$ at its centre.
 
Then, the definitions
(\protect\ref{e-defnxy}), (\protect\ref{e-defnphipi}) and
(\protect\ref{e-defns}) together with the values in 
eqns (\protect\ref{e-phieqval}), (\protect\ref{e-phieqpolval}) 
and (\protect\ref{e-sintheta}) give 
the vertex conversions
shown in Table~\ref{t-bcequat}. 

The requirements of
\begin{list}{(\roman{enumi})}{\usecounter{enumi}}
\item isolatitude pixel positions, and
\item equal pixel areas 
\end{list}
together with the vertex conversions in Table~\ref{t-bcequat} 
yield the linear transformations
\begin{eqnarray}
\phi_\pi &=& x /4 \nonumber \\
s &=& {{2 \over 3} (y-1) }.
\label{e-solneq}
\end{eqnarray}

\tbcpolar

\subsubsection{Polar regions: 
transformations from unit interval to $\phi,\theta$}
\label{s-fifjpol}

Without loss of generality, 
the polar region ``1'', as shown  
in Figs~\ref{f-twopithree},~\ref{f-twelveregions}, 
adjacent to the equatorial region labelled ``5'', is considered.

The vertex conversions for this polar region 
are those in Table~\ref{t-bcpolar}. 

Vertex conversion Eq.~(\ref{e-bcpolar}).F 
is a requirement of continuity of $\phi_\pi$ between
adjacent polar regions: the two boundaries of the upper half of the
region are the lines $x=y-2$ and $-x=y-2$ 
(respectively upper left and upper right boundaries of the unit 
interval in 
Fig.~\ref{f-fifj}) and for any fixed value of $y$, 
must cover the angular interval $[0,\pi/2]$, i.e. the $\phi_\pi$ value
at constant $y$ must increase by $\Delta \phi_\pi = 1/2$ between
these two limits. Hence, Eq.~(\ref{e-bcpolar}).F.

The region is illustrated in Fig.~\ref{f-fifj}.

The lower half of the `polar' region, where $y \le 1,$  is really 
in the equatorial part of the sphere, so a similar linear solution 
applies as above:
\begin{eqnarray}
\phi_\pi &=& {1 + x \over 4}  \nonumber \\ 
s &=& {2 y \over 3},
\label{e-yeq}
\end{eqnarray}
providing continuity and 
satisfying the isolatitude and equi-pixel requirements and the vertex
conversions 
(\ref{e-bcpolar}).B, 
(\ref{e-bcpolar}).C and 
(\ref{e-bcpolar}).D in Table~\ref{t-bcpolar}.

By continuity, (\ref{e-bcpolar}).C and (\ref{e-bcpolar}).D 
must also be satisfied for limiting cases in 
the upper half of the region as $y \rightarrow 1^+$. 

The isolatitude and equi-pixel requirements imply that for $y \ge 1$,
$\phi_\pi$ is linear in $x$, and in order to satisfy 
(\ref{e-bcpolar}).C and (\ref{e-bcpolar}).D, it follows that 
\begin{equation}
\phi_\pi = {1 + f(y) x \over 4}
\end{equation}
for some function $f$ which does not depend on $x$ and which 
satisfies 
\begin{equation}
f(y=1) = 1
\label{e-fyrequirement}
\end{equation}
in order to be continuous with the region $y \le 1$.

Boundary condition Eq.~(\ref{e-bcpolar}).F  (Table~\ref{t-bcpolar}) is 
\begin{equation}
\phi_\pi(-x=y-2) = \phi_\pi(x=y-2) + 1/2 
\end{equation}
\begin{equation}
\Rightarrow {1 + f(y) [-(y-2)] \over 4}
           = {1 + f(y) (y-2) \over 4} + 1/2 
\end{equation}
\begin{equation}
\Rightarrow   f(y) [-(y-2)] = f(y) (y-2) + 2 
\end{equation}
\begin{equation}
\Rightarrow   f(y) = {1 \over 2-y}.
\end{equation}

Because boundary condition F implicitly includes conditions C and D
when $y=1$, the required constraint in 
Eq.~(\ref{e-fyrequirement}) is satisfied, i.e. $f(1)=1$.

Thus, 
\begin{equation}
\phi_\pi = { 2 - y + x \over 4 (2-y)}.
\label{e-phipol}
\end{equation}

To obtain $s$, note that equality of pixel areas for $y \ge 1$ and
$y \le 1$ requires that
\begin{equation}
\left.{\partial s \over \partial y }\right|_{y \ge 1} 
\left.{\partial \phi_\pi \over \partial x }\right|_{y \ge 1} 
=
\left.{\partial s \over \partial y }\right|_{y \le 1} 
\left.{\partial \phi_\pi \over \partial x }\right|_{y \le 1} .
\end{equation}

Equations (\ref{e-phipol}), (\ref{e-yeq}) then give
\begin{equation}
\left.{\partial s \over \partial y }\right|_{y \ge 1} \; {1 \over 4 (2-y)} 
= {2 \over 3} \; {1 \over 4} 
\end{equation}
\begin{equation}
\Rightarrow 
\left.{\partial s \over \partial y }\right|_{y \ge 1} 
= {2 \over 3} (2-y).
\end{equation}

The isolatitude condition implies that
\begin{equation}
{\partial s \over \partial x } = 0, 
\end{equation}
so integration gives
\begin{equation}
s = \mbox{constant} - {1 \over 3} (2-y)^2, 
\end{equation}
and vertex conversion D provides the constant, so that
\begin{equation}
s = 1 - {1 \over 3} (2-y)^2.
\label{e-spol}
\end{equation}

Taking the limit of $\phi_\pi$ 
as $(x,y) \rightarrow (0,2)$, within the range 
$ y-2 \le x \le 2-y,$ clearly provides vertex conversion E.

\section{Conclusion}

The solution to the regionalisation of the sphere recommended by 
\nocite{Healpix99}{G\'orski} {et~al.} (1999b), with the three requirements
\begin{list}{(\roman{enumi})}{\usecounter{enumi}}
\item isolatitude, i.e. 
equal spacing of pixels in latitude, along lines of constant latitude,
\item equal pixel areas, and
\item hierarchical ordering of pixels within square unit intervals
which map to equal solid angles,
\end{list}
has been completed by deriving, in particular,
the angle $\theta^*$ separating polar
and equatorial regions, and the full transformations 
from the unit interval $[0,1] \wedge [0,1]$ 
to spherical coordinates in a polar region.

Together with the binary ordering presented by \nocite{Healpix99}{G\'orski} {et~al.} (1999b), 
the full solution is as follows.

Binary, bitwise mappings relate 
fractional positions $(f_i, f_j)$ 
within the $k$-th square unit interval to a pixel number.
With $n \ge 1$ binary digits, the pixel centred at $(f_i,f_j)$ 
with binary representations
\begin{equation}
f_i = 0.b_{2n-1}...b_3b_1 1, f_j= 0.b_{2n}...b_4b_2 1,
\end{equation}
where $b_i \in \{0,1\}$, 
in the $k$-th region shown in Fig.~\ref{f-twelveregions},
is pixel number
\begin{equation}
\left[ (k-1)\; 2^{2n} \right]_{10} + \left[b_{2n}...b_2b_1\right]_2,
\end{equation}
where the first term is shown in decimal notation and the second
in binary notation, as mentioned above in Eq.~(\ref{e-fullpixelnumber}).

The vertices along the equator of the equatorial regions are 
(from Eq.~(\ref{e-phieqval}))
\begin{equation}
\phi_\pi = 1/4,\; 3/4,\; 5/4,\; 7/4, \theta = 0 ;
\end{equation}
the vertices separating equatorial and polar regions are
(from Eq.~(\ref{e-phieqpolval}))
\begin{eqnarray}
&(\phi_\pi = 0,\; 1/2,\; 1,\; 3/2, \; \theta = \theta^*),&  \nonumber \\
&(\phi_\pi = 0,\; 1/2,\; 1,\; 3/2, \; \theta = -\theta^*) ;&  
\end{eqnarray}
and the  two polar vertices are 
(from Eq.~(\ref{e-phipolval}))
\begin{equation}
\phi_\pi \; \mbox{undefined},\; \theta = -\pi/2, +\pi/2.
\end{equation}

Mappings from $(x \equiv f_i - f_j, y \equiv f_i + f_j)$ to
$(\phi_\pi \equiv \phi/\pi, s \equiv \sin \theta)$, are found, without
loss of generality, for one equatorial region and one polar region:
\begin{eqnarray}
&&
\left.
\begin{array}{ccc}
\phi_\pi &=& {x \over 4}  \\
s &=& {2 \over 3} (y-1) 
\end{array}
\right\}
\mbox{\quad equatorial region}  \nonumber  \\ \nonumber  \\
&&
\left.
\begin{array}{ccc}
\phi_\pi &=& { 1 + x \over 4  }\nonumber \\ 
s &=& {2 y \over 3} 
\end{array}
\right\}
\mbox{\quad polar region}, \; y \le 1 \nonumber \\ \nonumber  \\
&&
\left.
\begin{array}{ccc}
\phi_\pi &=& \frac { 2 - y + x}{4 (2-y)} \\
s &=& 1 - {1 \over 3} (2-y)^2
\end{array}
\right\}
\mbox{\quad polar region}, \; y \ge 1
\label{e-fullsoln}
\end{eqnarray}
summarised here for convenience from 
equations (\ref{e-solneq}), (\ref{e-yeq}),
(\ref{e-phipol}) and (\ref{e-spol}).

\section*{Acknowledgments}

We thank Magdalena Cechowska for a careful reading of this paper.

\subm{ \clearpage }

\nice{

}


\end{document}